\documentstyle[pra,aps,twocolumn,epsfig]{revtex}

\begin{document}
\draft
\flushbottom
\twocolumn[
\hsize\textwidth\columnwidth\hsize\csname @twocolumnfalse\endcsname

\title{Vibrating soap films: An analog for quantum chaos on billiards}
\author{E. Arcos, G. B\'aez, P. A. Cuatl\'ayol, M. L. H. Prian, 
R. A. M\'endez-S\'anchez$^a$ and H. Hern\'andez-Salda\~na.}
\address{Laboratorio de Cuernavaca, Instituto de F\'{\i}sica, 
University of Mexico (UNAM) \\
A. P. 20-364, C. P. 01000, M\'exico D. F. MEXICO. \\
and \\
Facultad de Ciencias, UNAM,
University of Mexico (UNAM).
04510, M\'exico D. F. MEXICO. } 
\maketitle

\tightenlines
\widetext
\advance\leftskip by 57pt
\advance\rightskip by 57pt
\begin{abstract}
We present an experimental setup based on the normal modes of vibrating
soap films which shows quantum features of integrable and chaotic billiards.
In particular, we obtain the so-called {\it scars } --narrow linear
regions with high probability along classical periodic orbits-- for the
classically chaotic billiards. We show that these scars are also visible at
low frequencies. Finally, we suggest some applications of our experimental
setup in other related two-dimensional wave phenomena.
\end{abstract}
\vskip .5cm
]

\narrowtext
\tightenlines
\setcounter{equation}{0}

\section{INTRODUCTION}
\begin{figure}[tbp]
\vspace*{-.3in}
{\psfig{figure=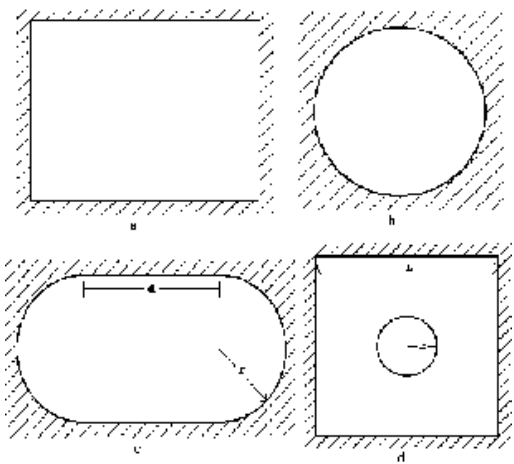,height=6.2cm}} 
\vspace*{.12in}
\caption{Integrable billiards: (a) rectangular and (b) circular. Chaotic
billiards: (c) the Bunimovich stadium, formed by two semicircumferences of
radius $r$ joined by two segments of length $d$; (d) the Sina\"{\i} billiard,
formed by a square with a circle of radius $r$ inside} \label {fig:1}
\end{figure}

\vspace*{-0.2in}
In recent years, there has been increasing interest in the properties of
quantum systems whose classical analogs are chaotic.\cite{qc,berry} Part of the
work in this new field, called quantum chaos,\cite{berry} refers to
essentially two-dimensional cavities or wells of infinite potential called
billiards. These billiards can take different forms, such as rectangles,
circles and other more complicated geo\-me\-tries (see Fig.~\ref{fig:1}). 
The circle
and the rectangle correspond to integrable systems.\cite{berry} Furthermore,
we include in Fig.~\ref{fig:1} 
the so--called Bunimovich stadium and the Sina\"{\i}
billiard which are completely chaotic.\cite{chaotic} However, intermediate
situations, i.e., systems with both integrable and chaotic behaviors, are
the most common type of dynamical systems.\cite{phystoday}

The quantum analog of a classical billiard is called a quantum billiard and
its eigenfunctions are closely related to the classical features of the
billiard.\cite{heller} Quantum billiards obey the Helmholtz equation with
vanishing amplitude on the border (homogeneous Dirichlet boundary
conditions). Such systems can be simulated as a drum or any other membrane
vibrating in a frame. The principal purpose of this paper is to show that
the quantum behavior of classically chaotic and integrable billiards can be
modeled in a classroom with an analog experiment. The experimental setup
basically contains a function generator and a mechanical vibrator and is
based on the vibrations of a soap film.\cite{both} Thus, as Feynman
said in 1963, ``the same equations have the same solutions''.\cite{feynman}

In the next section we briefly discuss classical and quantum billiards. In
Sec.~III we introduce our experimental setup and show the analogy
with the quantum billiard. Other uses of our experimental setup are
discussed in the same section. Some remarks are given in the conclusion.

\vspace*{-0.2in}
\section{CLASSICAL AND QUANTUM BILLIARDS}
\vspace*{-0.1in}

\begin{figure}[tbp]
{\psfig{figure=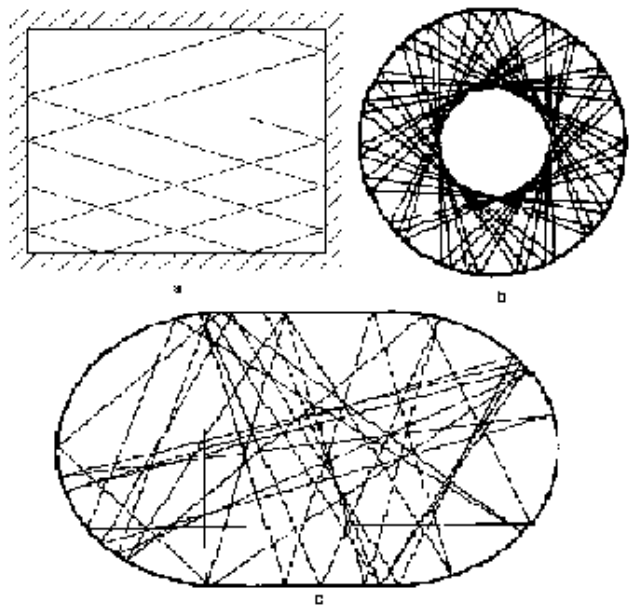,height=8cm}} 
\vspace*{.12in}
\caption{Typical trajectories in billiards: (a) in the rectangle, (b) in the
circle (This trajectory shows a caustic), and (c) in the Bunimovich stadium.}
\label {fig:2}
\end{figure}

\begin{figure}[tbp]
{\psfig{figure=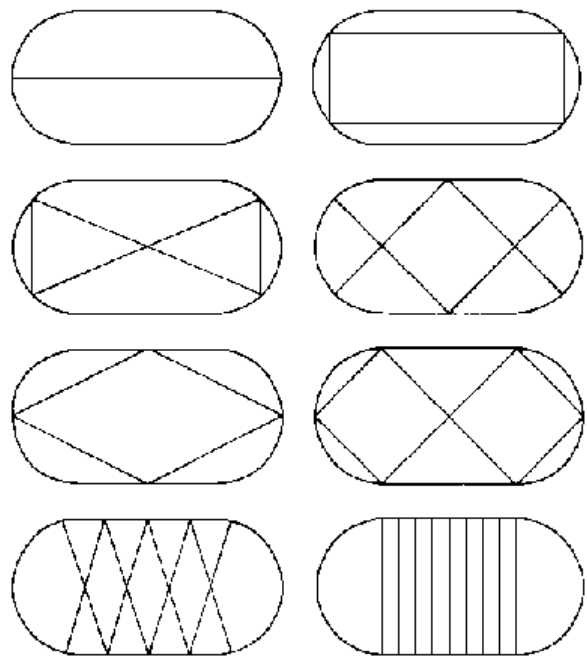,height=8cm,angle=90}} 
\vspace*{.12in}
\caption{Periodic orbits in the Bunimovich stadium. Notice that in some
cases more than one orbit is present and that all obey the reflection law.}
\label {fig:3}
\end{figure}

In order to make a more explicit definition of the classical billiard we take a
two-dimensional region denoted by $R$ and define the potential $V$ for the
particle as

\begin{equation}
V=\left\{ 
\begin{array}{cc}
{0} & {\rm in}\qquad {R} \\ 
{\rm \infty } & {\rm otherwise.}
\end{array}
\right. 
\end{equation}
This means that inside $R$ the particle is free and moves in straight lines.
When the particle collides with the boundary, it bounces following the
law of reflection. Under this dynamics, the rectangle and circle billiards are
re\-gular. Ty\-pic\-al trajectories inside are shown in Fig.~\ref{fig:2}. 
On the other
hand, the dynamics of a particle in the stadium as well as in the Sina\"{\i}
billiard, are chaotic. Almost all trajectories for these billiards are
ergodic and exponentially divergent. Roughly speaking, this is so because in
the Sina\"{\i} billiard, two very close trajectories are separated when one
collides (or both) with the central circle. After some time, the separation
between the trajectories (in phase space) is exponential. In the stadium
two particles with very close initial conditions are focused when they
collide with one semicircle. After this focusing, they began to separate
until they bounce again but now on the other semicircle. The exponential
divergence appears because the separation time is greater than the focusing
time. Apart from these trajectories there also exist periodic orbits. These
trajectories are unstable and typically isolated. Their number increases
exponentially as function of their length, but they are of measure zero in
phase space. We may find also families of unstable and
non-isolated periodic orbits such as the ``bouncing-ball'' orbits in which
the particle bounces between the two parallel segments of stadium. Some
periodic orbits for the Bunimovich stadium are shown in Fig.~\ref{fig:3}.

The time independent Schr\"odinger equation for the potential defined in
Eq.~(1) is

\begin{equation}
\begin{array}{r}
\nabla ^2\Psi +k^2\Psi =0 \\ 
\Psi =0
\end{array} \label{eq:2}
\begin{array}{cc}
{\rm in} & R, \\ 
{\rm on\ the\ boundary\ of} & R,
\end{array}
\end{equation}
with the wave number $k=({2mE/\hbar )^{1/2}}$. Here, $E$ and $m$ are the
energy and mass of the particle and $\hbar $ is the Planck constant. The
homogeneous Dirichlet boundary condition is obtained because if $V=\infty $
the wave function vanishes. The Helmholtz equation in $R$ and the Dirichlet
boundary condition define the quantum billiard. Note that this is just the
equation for the normal modes of a membrane if we interpret the functions as
vibration amplitudes.

\begin{figure}[tbp]
{\hspace*{-.5cm}\psfig{figure=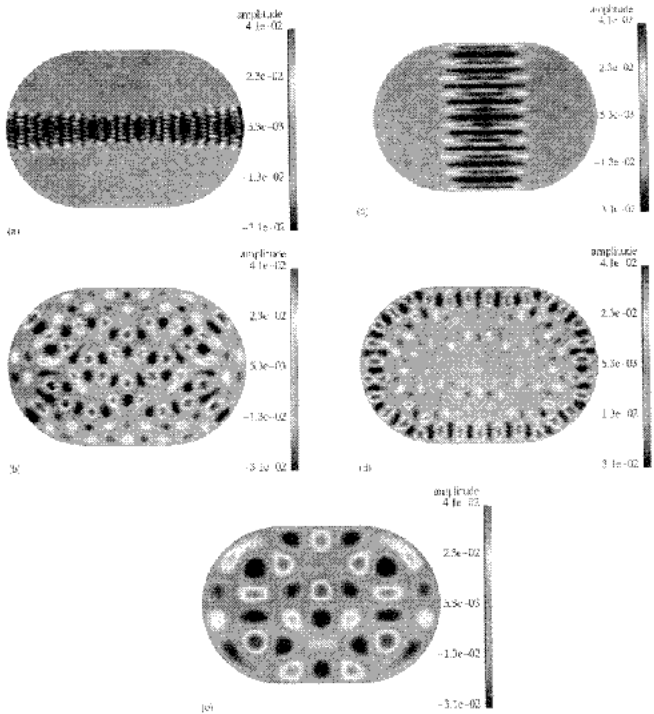,height=10cm}} 
\vspace*{.12in}
\caption{Eigenfunctions of the Bunimovich stadium calculated by the finite
element method for a quadrant of stadium with Dirichlet boundary
conditions. The maxima and minima of each normal mode correspond to the
darkest zones; (a) a typical state scarred by a short classical orbit
connecting the two extremes of the stadium, (b) a state scarred by an orbit of
large period, (c) a typical state scarred by a ``bouncing-ball'' orbit, (d)
``whispering gallery'' state, and (e) a typical noisy state if 
one looks only at a quadrant.} \label {fig:4}
\end{figure}

For classically integrable billiards, the eigenfunctions are well-known. For
example the solutions for circular and rectangular billiards are Bessel
functions and sinusoidal functions, respectively.\cite{sines} On the other
hand, the features of wave functions for classically chaotic billiards have
been well studied numerically by Heller.\cite{heller} Recently, experiments
in microwave cavities have been per\-for\-med.\cite{sridhar} In 
Fig.~\ref{fig:4} we
show eigenfunctions of the Bunimovich stadium we calculated numerically
using the finite element method.\cite{mori} However they can alternatively
be calculated by standard software. The eigenfunctions of fi\-gu\-res 
\ref{fig:4}(a)-(c) show certain similarities with the orbits of 
Fig.~\ref{fig:3}. Following Heller, we say that the eigenfunctions are 
``scarred'' by the orbits. Fig.~\ref{fig:4}(d) shows what is
called a ``whispering gallery'' state, because there exist certain galleries%
\cite{gallery} in which the sound travels inside them, following the border.
This kind of state is associated with orbits also close to the boundary. The
eigenfunction shown on Fig.~\ref{fig:4}(e) resembles noise\cite{berry} when 
we see only a quarter of stadium.

The appearance of the scars is quite well understood\cite{bogomolny} based
on the theo\-retical work by Selberg, Gutzwiller and Balian.\cite{trace
formula} Due to the low density of the short periodic orbits they may well
be seen in quantum experiments and simulations either as dominant features
in the Fourier spectrum or as scars. While the Fourier analysis of
experimental data in atomic\cite{atoms} and molecular physics\cite{molecules}
is quite striking, direct demonstrations of scars are difficult even in
microwave cavities. A simple explanation of scarring is based on de Broglie
waves. Close to the periodic orbit there exist standing de Broglie waves
whose wavelength $\lambda $ is associated to the length $L$ of the periodic
orbit: 
\begin{equation}
2L=n\lambda ,\qquad n=1,2,3,\dots .
\end{equation}
These de Broglie waves are localized around periodic orbits due to the
exponential divergence of nearing trajectories. In the next section we will
show how a simple demonstration setup can display the most interesting
features of the eigenfunctions on a soap film.

\vspace*{-0.2in}
\section{Soap Film Analogy}
\vspace*{-0.1in}
The normal modes of a rectangular and circular soap film have been well
studied.\cite{cyril,bubbles ajp,bergmann} 
A textbook showing these eigenfunctions is French's book entitled {\it %
Vibrations and Waves.}\cite{french} The governing equation is the time
independent wave equation (Eq.\ref{eq:2}), but now with a different 
interpretation: $\Psi $ is the membrane vibration amplitude and the wave 
number is now $k=\omega / v$, with $\omega $ the angular frequency and 
$v$ the speed of the transverse waves on the membrane.

\begin{figure}[tbp]
{\hspace*{-.5cm}\psfig{figure=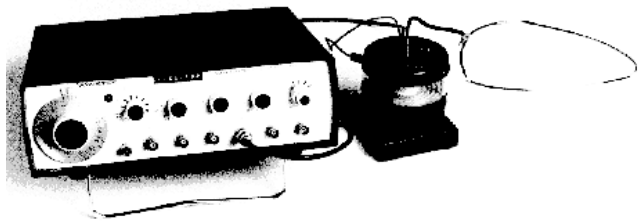,height=3.4cm}} 
\vspace*{.12in}
\caption{ Experimental setup used to show the quantum features of chaos.
Notice that the membrane is excited in a way that breaks the symmetry.} 
\label{fig:5}
\end{figure}

A problem arises for the experimental setup of this analog: At high
frequencies (corresponding to the semiclassical limit), the damping is large.
To solve this problem we feed energy into the system permanently with an
external re\-so\-na\-tor of a well-defined but variable frequency. We chose
to feed the external frequency into the system by vibrating the wire that
delimits our soap film. We use a mechanical vibrator (PASCO Scientific model
SF-9324, see Fig.~\ref{fig:5}) connected to a function generator (we used a 
generator Wavetek model 180\cite{other uses}) and used wires with different 
shapes (rectangle, circle, Bunimovich stadium, Sina\"{\i} billiard, and some 
other of interest\cite{hans}) Alternatively a speaker could be used to transmit
the frequency through the air.\cite{bubbles ajp,bergmann,walker} 
The chemical formula for a soap film with large duration is
given by Walker,\cite{walker} but can be made up with soap, water and
glycerin by trial and error.

\begin{figure}[tbp]
{\hspace*{-.5cm}\psfig{figure=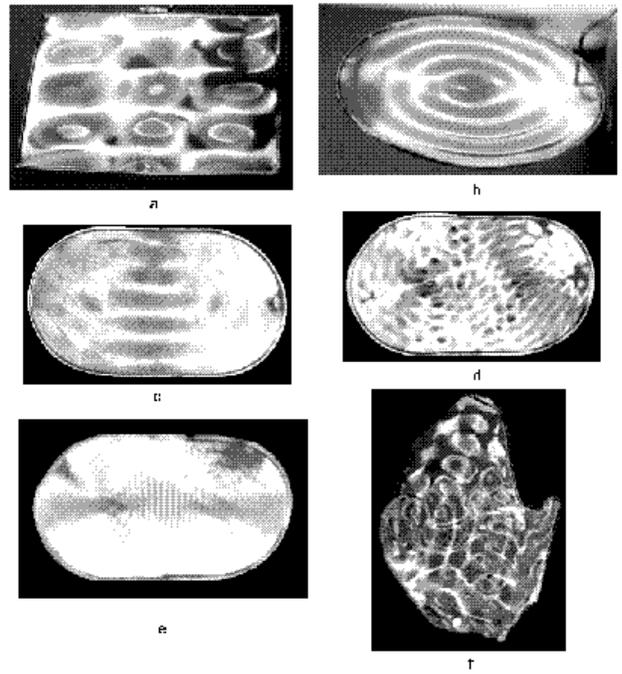,height=9cm,angle=0}}
\vspace*{.12in}
\caption{Normal modes for different shapes. (a) Low mode for the
rectangle at a frequency of $22.48$ Hz (The frequency changes slightly for
different soap solutions and different thicknesses of the soap film.) The
rectangle is of size $0.215$ m$\times 0.155$ m. (b) Bessel function for a
circular-shape wire. The radius of the wire is $0.092$  and the excitation
frequency is around $33.5$ Hz. (c) A ``bouncing-ball state'' near to $40$ Hz.
(d) A scarred state corresponding to the numerical one showed in
Fig.~\ref{fig:4}(b) at frequency close to $110$ Hz. (e) A scarred state by a
short classical orbit
connecting the two extremes of the stadium for a Bunimovich stadium with
$r=0.035\ $ and $d=0.035\ $. The corresponding frequency is about $478$ Hz.
(f) A normal mode for a wire with ``old Tenochtitlan lake'' form at 
frequency $\nu =48.88$ Hz.} 
\label {fig:6}
\end{figure}

We can now start the demonstration. In Fig.~\ref{fig:6}(a) we show a normal 
mode of the rectangle, in agreement with the known result. 
The Fig.~\ref{fig:6}(b) shows a normal
mode for a circle displaying a Bessel function. Roughly speaking, the
shining and dark zones establish a periodic pattern associated to the normal
mode. Although the pattern established cannot give a quantitative measure of
the amplitude, it is sufficient to give an idea of the form of the normal
mode and to identify it. However, the more interesting normal modes are some
of the classically chaotic billiards. In Fig.~\ref{fig:6}(c) we show a
``bouncing--ball'' state at low energies for a Bunimovich stadium. The
alternating dark and shining zones in this and in the following figures
make evident the presence of standing waves in the membrane. These waves are
associated to de Broglie waves in the quantum billiard and at the same time
they are associated with periodic orbits in the classical billiard. In
Figs.~\ref{fig:6}(d) and (e) scarred eigenfunctions are displayed, 
the latter in the high-frequency regime. In order to show that they are 
scars and not some spurious
effect of our very simple experimental setup, we calculate numerically the
normal modes at frequencies near the experimental ones. We observe in 
Fig.~\ref{fig:4}
some of these eigenfunctions and the corresponding orbits in Fig.~\ref{fig:3}.

We want to mention that the normal modes in soap films can also be used for
other two-dimensional wave phenomena, such as the search of normal modes of
the clay layer corresponding to the old Tenochtitlan lake, which plays a
crucial role in the earthquake damage patterns of Mexico City.\cite{nature}
The application in this case comes from the fact that the upper clay layer
of the Mexico Valley, as well as the soap films, are practically
two-dimensional.\cite{nature} A wire may readily be shaped to the
corresponding boundary and the results are shown in Fig.~\ref{fig:6}(f).

\begin{figure}[tbp]
{\hspace*{-.5cm}\psfig{figure=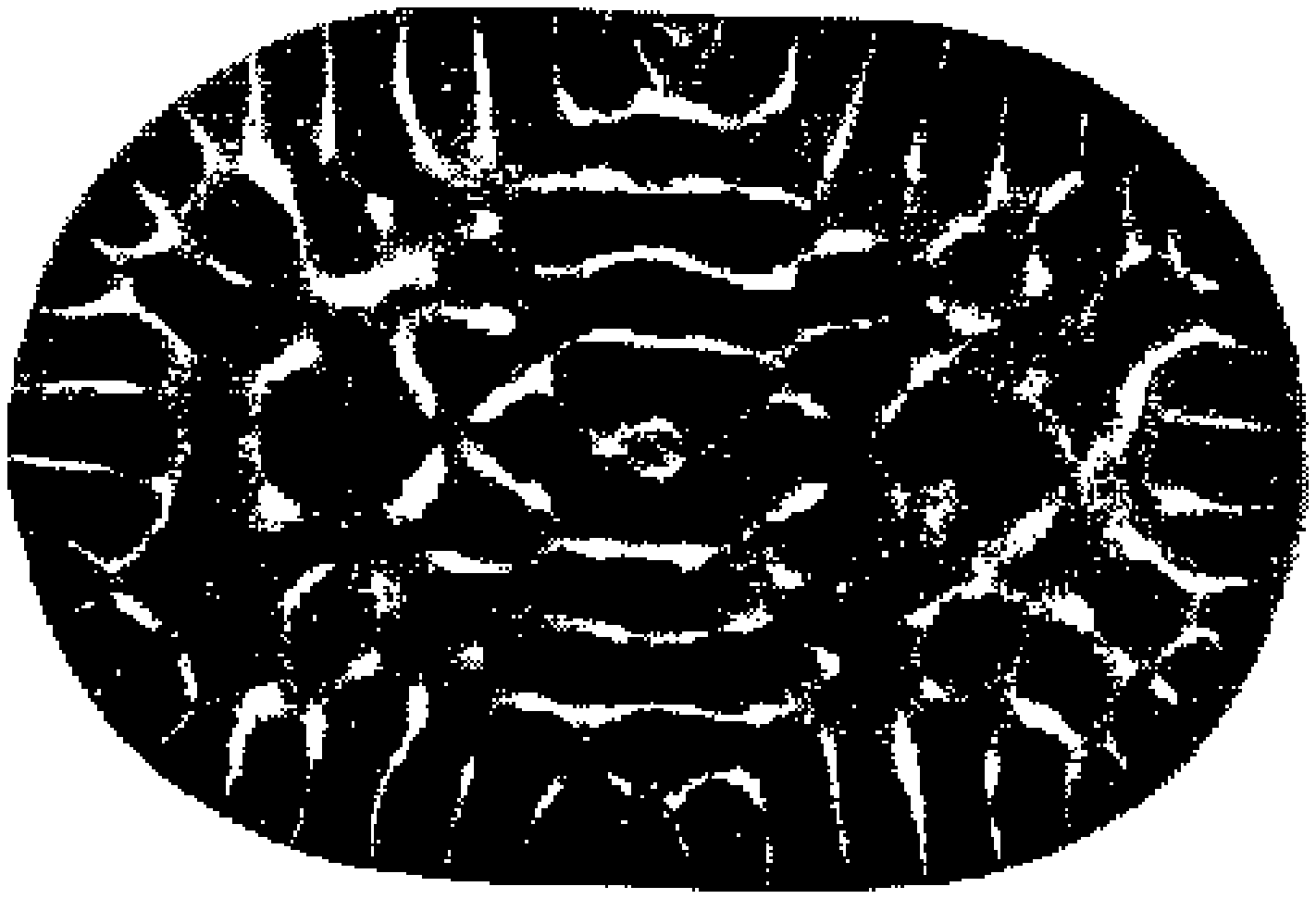,height=2.9cm}} 
{\psfig{figure=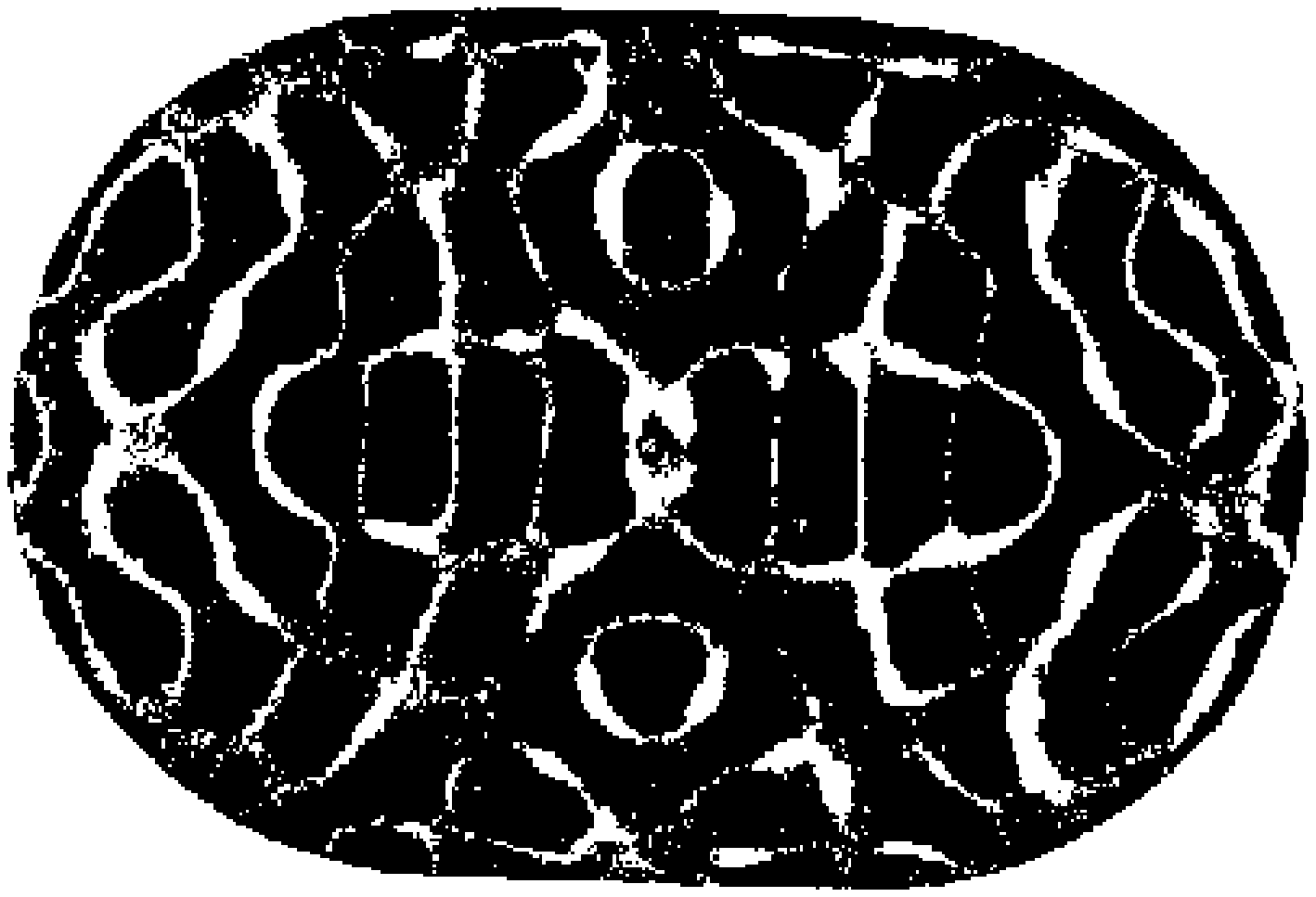,height=2.9cm}}
\vspace*{.12in}
\caption{Nodal patterns observed on iron plates with stadium shape. The
plate was of $1$ mm thickness and we used $r=0.1$m and $d=0.1$ m. (a)A
whispering gallery at $5.79$kHz. Notice that this pattern is better defined
near the boundary and between the two segments of the stadium. This means
that the amplitudes are higher in these regions. (b) A scar which reflects the
orbit connecting the two extremes of the stadium. In this case the pattern
is better defined on the periodic orbit. The frequency for this case was
around $4.22$ kHz.} \label {fig:7}
\end{figure}

Moreover, the experimental setup which we present may be used to show other
related wave phenomena. Typical examples are scars on li\-quids,\cite{agua}
Faraday waves --crystallographic patterns in large-amplitude waves\cite
{faraday}--, cuasi-crystalline patterns on liquids,\cite{fouve} sand dynamics%
\cite{sand} or normal modes of Chlad\-ni's plates.\cite{placas} These can be
done by replacing the wire in our experimental setup with water containers or
by thin plates. If we want to see scars on surface waves, we must put a
stadium-shaped tank filled by water. Faraday waves can be obtained adding
shampoo to the water and increasing the amplitude and frequency but
decreasing the level of the liquid up to several millimeters. It is not
necessary to use tanks with different shapes because the patterns do not
depend on the boundary. If we want to observe quasi-crystalline patterns
with this experimental setup, we must excite the mechanical vibrator with--
at least --two frequencies.\cite{fouve} This is easily obtained by changing
the sinusoidal time--dependence of the of the driving force to a triangular
one.\cite{triangular} Another application is the demonstration of the modes
of thin plates. In this case, as well as in the water tanks, the whispering
gallery states are easily visible. We may use ellipses, pentagons or any
other shape. All these plates are excited in a point in which the mechanical
vibrator loads them. Figure \ref{fig:7}(a) shows a whispering gallery state 
for a stadium-shaped iron plate. 
Fig.~\ref{fig:7}(b) shows a scarred pattern for the same 
plate.
Finally, if we consider a tank with a layer of sand of variable thickness,
we may study several topics on dynamics of granular media. As an example we
can study ``standing waves'' on sand.\cite{sand}

\vspace*{-0.2in}
\section{Conclusions}
\vspace*{-0.1in}

The analog model of soap films for the quantum billiard gives a
de\-mons\-tra\-tion of quantum chaos features. For integrable regions the
normal modes correspond to the eigenfunctions of two-dimensional square box
and circle. For classically chaotic billiards the normal modes correspond to
scarred eigenfunctions in the semiclassical limit. The experimental setup
presented here is particularly cheap, simple and elegant-- i.e. the
alternative expe\-ri\-ments do not satisfy the Helmholtz equation and/or
boundary condition. Microwave cavities are expensive and do not display the
scars in a directly visible ma\-nner--. Thus the simplicity of the
experimental setup and the facility to put any shape, makes it very suitable
for the undergraduate laboratory. Furthermore the normal modes of soap
films can be used to de\-mons\-trate other two-dimensional analog phenomena.
Finally, we want to mention that the proposed experimental setup can be
quickly changed to study other related wave phenomena. A wide variety of
highly nontrivial wave-like phenomena can be displayed with minimal
experimental requirements.

\vspace*{-0.2in}
\section{Acknowledgments}
\vspace*{-0.1in}

We want to thank to T. H. Seligman, F. Leyvraz and J. A Heras, for their
valuable comments. Also we would like to thank to the C. O. F., F\'{\i}sica
General and F\'{\i}sica Mo\-der\-na Laboratories of the Facultad de Ciencias
U.N.A.M. and specially to Felipe Ch\'{a}vez. Finally we thank L. M. de la
Cruz for his useful help with some of the figures. The numerical work was done
in the Cray supercomputer of U.N.A.M.

\begin{enumerate}
\item[$^{a)}$]  E-mail: mendez@ce.ifisicam.unam.mx
\vspace*{-.28in}
\end{enumerate}

\begin{thebibliography}{99}
\bibitem{both}  Both the Schr\"odinger equation for a billiard and the
wave equation for a membrane are reduced to the Helmholtz equation when the
time independent part is taken. Nevertheless, to deduce the wave equation
for the membrane air, gravity and the damping in
the high frequency regime are usually neglected.
\end{thebibliography}
\end{document}